\documentstyle[12pt,epsf]{article}
\baselineskip=\normalbaselineskip
\ifx\TwoupWrites\UnDeFiNeD\else\target{\magstepminus1}{11.3in}{8.27in}
	\source{\magstep0}{7.5in}{11.69in}\fi
\newfont{\fourteencp}{cmcsc10 scaled\magstep2}
\newfont{\titlefont}{cmbx10 scaled\magstep2}
\newfont{\authorfont}{cmcsc10 scaled\magstep1}
\newfont{\fourteenmib}{cmmib10 scaled\magstep2}
	\skewchar\fourteenmib='177
\newfont{\elevenmib}{cmmib10 scaled\magstephalf}
	\skewchar\elevenmib='177
\makeatletter
\newcommand\nonsequentialeqnum{
	\@addtoreset{equation}{section}
	\def\theequation{\arabic{section}.\arabic{equation}}}
\newif\ifp@bblock  \p@bblocktrue
\newcommand\nopubblock{\p@bblockfalse}
\newcommand\topspace{\hrule height 0pt depth 0pt \vskip}
\newcommand\p@bblock{\begingroup \tabskip=\hsize minus \hsize
	\baselineskip=1.5\ht\strutbox \topspace-2\baselineskip
	\halign to\hsize{\strut ##\hfil\tabskip=0pt\crcr
	\the\Pubnum\crcr
	KUCP-0065\crcr
	\the\date\crcr}\endgroup}
\newcommand\YITPmark{\hbox{\fourteenmib YITP\hskip0.2cm
        \elevenmib Uji\hskip0.15cm Research\hskip0.15cm Center\hfill}}
\renewcommand\titlepage{\ifx\TwoupWrites\UnDeFiNeD\null\vspace{-1.7cm}\fi
	\YITPmark\vskip0.6cm
	\ifp@bblock\p@bblock \else\hrule height 0pt \relax \fi}
\makeatother
\newtoks\date
\newtoks\Pubnum
\newtoks\pubnum
\Pubnum={YITP/U-\the\pubnum}
\date={\today}
\newcommand{\frontpageskip}{\vspace{12pt plus .5fil minus 2pt}}
\renewcommand{\title}[1]{\frontpageskip
	\begin{center}{\titlefont #1}\end{center}\par}
\renewcommand{\author}[1]{\frontpageskip\par\begin{center}
	{\authorfont #1}\end{center}
	\nobreak
	}

\newcommand{\address}[1]{\par\begin{center}{\sl #1}\end{center}\par}
\newcommand{\andaddress}{\par\centerline{\sl and}\address}
\renewcommand{\thanks}[1]{\footnote{#1}}
\renewcommand{\abstract}{\par\frontpageskip\centerline{\fourteencp Abstract}
	\vspace{8pt plus 3pt minus 3pt}}

\def\mean#1{\overline{#1}}
\def\diff#1{\left[#1\right]}

\def\threeD{\,{}^3\!}
\def\para{{\scriptscriptstyle/\!/}}
\def\orth{{\scriptscriptstyle\perp}}
\def\odd{{(\r{o})}}
\def\oddv{{(\r{o}1)}}
\def\oddt{{(\r{o}2)}}
\def\even{{(\r{e})}}
\def\evens{{(\r{e}0)}}
\def\evenv{{(\r{e}1)}}
\def\event{{(\r{e}2)}}
\def\ovu#1#2{{#1}_\oddv^{#2}}
\def\ovd#1#2{{#1}_{\oddv #2}}
\def\ot#1{{#1}_\oddt}
\def\evu#1#2{{#1}_\evenv^{#2}}
\def\evd#1#2{{#1}_{\evenv #2}}
\def\es#1{{#1}_\evens}
\def\et#1{{#1}_\event}
\def\opot{\Phi_\odd}
\def\epot{\Phi_\even}

\catcode`@=11

\@addtoreset{equation}{section}   
\def\theequation{\arabic{section}.\arabic{equation}}

\def\appendix{\par
 \setcounter{section}{0}
 \setcounter{subsection}{0}
 \def\thesection{Appendix \Alph{section}}
 \def\thesubsection{\Alph{section}.\arabic{subsection}}
 \def\theequation{\Alph{section}.\arabic{equation}}}

\def\thebibliography#1{\section*{References\@mkboth
 {REFERENCES}{REFERENCES}}\list
 {\leftbibmark\arabic{enumi}\rightbibmark}{
 \settowidth\labelwidth{\leftbibmark #1\rightbibmark}\leftmargin\labelwidth
 \advance\leftmargin\labelsep
 \usecounter{enumi}}
 \def\newblock{\hskip .11em plus .33em minus -.07em}
 \sloppy\clubpenalty4000\widowpenalty4000
 \sfcode`\.=1000\relax}

\def\@cite#1#2{\leftcitemark{#1\if@tempswa , #2\fi}\rightcitemark}
\def\leftcitemark{[}
\def\rightcitemark{]}
\def\leftbibmark{[}
\def\rightbibmark{]}

\catcode`@=10

\def\Beq{\begin{equation}}
\def\Eeq{\end{equation}}
\def\Beqar{\begin{eqnarray}}
\def\Eeqar{\end{eqnarray}}
\def\Beqarn{\begin{eqnarray*}}
\def\Eeqarn{\end{eqnarray*}}

\def\linebreak{\hfill\break}
\def\r#1{{\rm #1}}

\def\bg#1{\hbox{\rlap{\hbox{$#1$}}\kern0.6pt $#1$}}
\def\Frac(#1/#2){\left(\frac{#1}{#2}\right)}

\def\therefore{\mbox{\setbox0=\hbox{X}\hbox{$\ldotp$}\raise0.7\ht0\hbox{$\ldotp$}\hbox{$\ldotp$}} }
\def\because{\mbox{\setbox0=\hbox{X}\raise0.7\ht0\hbox{$\ldotp$}\hbox{$\ldotp$}\raise0.7\ht0\hbox{$\ldotp$}}\kern0pt }

\def\CITE#1{$^{\hbox{\small \cite{#1}}}$}
\def\hannko{\rlap{\large $\bigcirc$}\kern2pt\hbox{\small $B0u(B}}

\def\I{{\cal I}}

\def\M{{\cal M}}
\def\N{{\cal N}}

\begin{document}
\thispagestyle{empty}
%
\pubnum{94-01}
\date{January 1994}
\titlepage

\title{
Does a domain wall emit gravitational waves?  \\
-- General-relativistic perturbative treatment --
}

\author{Hideo Kodama$^*$,
Hideki Ishihara$^{**}$
and Yoshihisa Fujiwara$^*$%
\footnote[2]{JSPS follow.}
}
\address{${}^*$ Uji Research Center, Yukawa Institute for Theoretical
Physics,\\
Kyoto University, Uji 611, Japan}
\andaddress{${}^{**}$ Department of Fundamental Sciences, FIHS, \\
Kyoto University, Kyoto 606, Japan}

\abstract{
The behavior of gravitational wave perturbations on a locally
Minkowskian spacetime background containing a planar domain wall is
investigated in the gauge-invariant general relativistic framework.
It is shown that for this particular background the domain wall does
not emit gravitational waves spontaneously by its free oscillation in
the first order, although it scatters incidental gravitational waves.
}

\newpage

\section{Introduction}

Domain walls are the simplest type of topological defects associated
with the symmetry breaking in unified
theories\CITE{Kibble.T1980,Vilenkin.A1985}.  Though stable domain wall
networks are not allowed in the early universe because they
overdominate the radiation energy density, unstable ones such as
domain wall bubbles and walls bounded by strings are expected to be
produced by cosmological phase transitions.  According to the work by
Vachaspati, Everett and
Vilenkin\CITE{Vachaspati.T&Everett&Vilenkin1984}, the dominant decay
process of domain walls is the emission of gravitational waves
provided that the energy loss rate is given by the quadrupole formula.
This result implies that the production of domain walls and their
decay may leave gravitational waves which are detectable by the
gravitational wave detectors to be constructed in the near future,
provided that the phase transition occurs at a sufficiently later
time, say around the time when the cosmic temperature is $10^6$GeV.
Though this is a quite fascinating prediction, it is not clear whether
it can be taken serious or not.  It is because there exists no
guarantee that the quadrupole formula does give a correct estimate on
the gravitational wave emission rate by topological defects which have
a large negative internal pressure.

In order to get some insights into this problem, in the present paper,
we investigate in the general relativistic framework the behavior of
perturbations of a spacetime which contains a domain wall.  As the
background spacetime we only consider the exact solution discussed by
Vilenkin\CITE{Vilenkin.A1983a} and Ipser and
Sikivie\CITE{Ipser.J&Sikivie1984} which is spherically symmetric and
locally Minkowskian outside a domain wall.  The domain wall is treated
as an infinitesimally thin singular hypersurface with a surface
energy-momentum density specific to the domain wall.  The whole
spacetime consists of two identical Minkowskian regions matched by the
domain wall.

The reason why we adopt this background spacetime is purely technical.
As systematically discussed by Gerlach and
Sengupta\CITE{Gerlach.U&Sengupta1979,Gerlach.U&Sengupta1979a,%
Gerlach.U&Sengupta1979b,Gerlach.U&Sengupta1980}, the spacetime
perturbations on a spherically symmetric background can be described
by gauge-invariant quantities which are local combinations of the
tensor harmonic expansion coefficients of the metric perturbations,
and the Einstein equations can be written in terms of them.  These
Einstein equations can be further reduced into a single master
equation for a single gauge-invariant variable corresponding to the
Zerilli variable\CITE{Regge.T&Wheeler1957,Zerilli.F1970}.
In our special background this
master equation can be easily solved explicitly in each Minkowskian
region.  The global solutions are constructed by matching these
solutions along the domain wall so that they satisfy the junction
condition of Israel\CITE{Israel.W1966}.

The organization of the paper is as follows.  In the next section we
describe the details of the background geometry.  Then in section 3 we
introduce the gauge-invariant variables, reduce the Einstein equations
written in terms of them into a single master equation, and give the
general solution to them.  In section 4, after rewriting the junction
condition as the boundary conditions on the master variables, we
construct the global solutions with the help of them and investigate
their structure. Section 5 is devoted to summary and discussion.

Throughout the paper we use the units $c=1$ and denote $8\pi G$ by
$\kappa^2$. The signature of Lorentzian metrics is chosen to be
$(-,+,+,\cdots)$.

\section{Background Geometry}

The background geometry we consider in this paper consists of two
spacetimes $(\M_\pm, g_\pm)$, both of which are some regions of the
Minkowski spacetime, connected by a three dimensional time-like
hypersurface $\Sigma$ representing a thin domain wall. We assume that
the whole spacetime manifold $\M=\M_+\cup\Sigma\cup \M_-$ is smooth and
$\Sigma$ is a smooth submanifold of $\M$.  We further require that the
spacetime metric $g_{\mu\nu}$ is $C^0$ in $\M$ and spherically symmetric.

The metric of a spherically symmetric spacetime is always written as
\Beq
ds^2 = \gamma_{ab}dy^a dy^b + r^2 \Omega_{pq}dz^pdz^q,
\Eeq
where $\Omega_{pq}dz^pdz^q$ is the standard metric of a unit Euclidean
2-sphere and $\gamma_{ab}$ and $r$ are functions of $y^a$. The
coordinates $y^p$ parametrize the two-dimensional orbit space $\N$
each point of which corresponds to a symmetry 2-sphere, and
$\gamma_{pq}dy^p dy^q$ gives the metric of this orbit space.
Throughout this paper the indices $a, \ldots, d$ run over 0 and 1, and
$p,\ldots, s$ over 2 and 3.

Let $D_a$ and $\hat D_p$ be the covariant derivatives with respect to
$\gamma_{ab}$ and $\Omega_{pq}$.  Then the covariant derivative
$\nabla_\mu$ with respect to the four-dimensional metric $g_{\mu\nu}$ is
expressed in terms of $D_a$ and $\hat D_p$.  For example, for a
contravariant vector $(V^\mu)=(V^a,V^p)$, we get
\Beqar
&&\nabla_a V^b = D_a V^b, \quad \nabla_a V^p = {1\over r}D_a(rV^p),
\label{NablaByD1}\\
&&\nabla_p V^a = \hat D_p V^a-{1\over r}D^a r V_p, \quad
\nabla_p V^q = \hat D_p V^q + {1\over r}D_a r V^a \delta^q_p.
\label{NablaByD2}
\Eeqar
Here $V^a$ and $V^p$ behave as scalars with respect to $\hat D$ and $D$,
respectively. The corresponding formulas for other tensors are easily
obtained from these with the help of the standard properties of the
covariant derivatives.

The domain wall is described by a time-like curve $\Gamma$ in the
orbit space $\N$. Since the spacetime is flat except on $\Sigma$, the
orbit space metric is also flat on $\N-\Gamma$. Further $r(y)$
satisfies the equations
\Beqar
&& D_aD_b r = 0,\\
&& D_ar D^ar =1.
\Eeqar

The whole spacetime $(\M,g)$ is constructed by gluing two regions
$\M_\pm$ in the four-dimensional Minkowski spacetime along boundary
hypersurfaces $\Sigma_\pm=\partial\M_\pm$. The possible boundary
surfaces are restricted by the junction conditions if we assume that
our spacetime is a limit of some family of smooth spacetimes.

Let $(x^\mu)$ be a coordinate system around $\Sigma$ in the whole
spacetime $M$ such that $x^p(p=2,3)$ coincide with the angular
coordinates $z^p$, and $x^j(j=1,2,3)$ gives a local coordinate system
of $\Sigma$. Further let $(q_{jk})_\pm$ be the intrinsic metric of
$\Sigma_\pm$ induced from $(g_{\mu\nu})_\pm$, $(n^\mu)_\pm$ the unit
normal vector to $\Sigma_\pm$ directed from $\M_-$ to $\M_+$, and
$(K^j_k)_\pm$ the extrinsic curvature of $\Sigma_\pm$ with respect to
$(n^\mu)_\pm$.  Then, as shown by Israel, the junction conditions are
given by the following equations:
\Beqar
&& \diff{q_{jk}}=0, \label{GJC1}\\
&& \diff{K^j_k}=\kappa^2 \left(S^j_k - {1\over2}\delta^j_k
S^l_l\right), \label{GJC2}
\Eeqar
where in general, for fields $Q_\pm$ on $\Sigma_\pm$, $\diff{Q}$
represents the difference
\Beq
\diff{Q} := Q_+ - Q_-,
\Eeq
and $S^j_k$ is the energy-momentum surface density of the domain wall.
The first condition simply states that $(q_{jk})_\pm$ coincides with
the intrinsic metric $q_{jk}$ of $\Sigma$ induced from $g_{\mu\nu}$.

In this paper we assume that $S^j_k$ is invariant under the local
three-dimensional Lorentz transformation in the tangent space of
$\Sigma$.  Hence $S^j_k$ can be expressed in terms of a positive
scalar field $\lambda$ as
\Beq
\kappa^2 S^j_k = -4 \lambda \delta^j_k, \label{EnergyMomentum:DW}
\Eeq
and constraint (\ref{GJC2}) is written as
\Beq
\diff{K^j_k} = 2\lambda \delta^j_k. \label{GJC2:DW}
\Eeq
{}From the momentum constraint on $\Sigma_\pm$, $\diff{K^j_k}$ satisfies
\Beq
\threeD\nabla_k\diff{K^k_j}-\threeD\nabla_j\diff{K} =0,
\Eeq
where $\threeD\nabla$ is the covariant derivative with respect to the
3-metric $q_{jk}$ of $\Sigma$.  This leads to the constraint
$\threeD\nabla_j \lambda=0$.  Hence $\lambda$ must actually be a
positive constant.

The above junction conditions uniquely specify $\Sigma_\pm$ to be
represented by the same time-like hyperboloid in the Minkowskian
spacetime.  In particular $(\M_+,g_+)$ and $(\M_-,g_-)$ must be two
copies of the region inside the hyperboloid.  To see this, let us
calculate $K^j_k$ explicitly.

First note that the unit normal vector $n^\mu$ has vanishing
$z^p$ components, hence can be regarded as a vector $n^a$ in $\N$. Let
$\tau^a$ be the future-directed unit tangent vector to the curve
$\Gamma$, and $\tau^\mu$ its counterpart in $\M$.  Then the projection
tensor to $\Sigma$ defined by
\Beq
q_{\mu\nu} := g_{\mu\nu} - n_\mu n_\nu
\Eeq
is expressed as
\Beq
q_{ab} = -\tau_a \tau_b, \quad q_{pq} = r^2 \Omega_{pq}, \quad q_{ap}=0.
\Eeq

In terms of these quantities the extrinsic curvature tensor
\Beq
K_{\mu\nu} := - q_\mu^\lambda \nabla_\lambda n_\nu
\Eeq
is expressed as
\Beqar
&& K_{ab} = \tau^c D_\para n_c q_{ab},\\
&& K_{ap} = 0,\\
&& K_{pq} = -r D_\orth r\Omega_{pq},
\Eeqar
where
\Beq
D_\orth := n^a D_a, \quad D_\para := \tau^a D_a.
\Eeq
Hence junction condition (\ref{GJC2:DW}) is written as
\Beqar
&& \left(\tau^c D_\para n_c\right)_\pm = \pm \lambda + \mu_1, \label{DWeq1}\\
&& \left({1\over r}D_\orth r\right)_\pm = \mp \lambda + \mu_2, \label{DWeq2}
\Eeqar
with some spherically symmetric functions $\mu_1$ and $\mu_2$ on $\Sigma$.

$\mu_1$ and $\mu_2$ cannot be arbitrary.  First from the Hamiltonian
constraint
\Beq
\threeD R + K^j_k K^k_j - K^2 =0,
\Eeq
it follows that
\Beq
\mean{K^j_k}\diff{K^k_j} - \mean{K}\diff{K} = 0, \label{diffR}
\Eeq
where $\mean{Q}$ for $Q_\pm$ on $\Sigma_\pm$ denotes the mean value
\Beq
\mean{Q} := {1\over2}(Q_+ + Q_-).
\Eeq
Inserting the above expressions of $K^j_k$ into this equation we get
$\mu_1=\mu_2$.

Next in the coordinates $(t_\pm,r)$ of $\N_\pm$ for which
\Beq
(\gamma_{ab})_\pm dy^ady^b = -dt_\pm^2 + dr^2,
\Eeq
from Eq.(\ref{DWeq2}) $n^a$ is written as
\Beq
(n^t)_\pm = \epsilon \sqrt{(n^r)_\pm^2 -1}, \quad (n^r)_\pm=(\mp \lambda +
\mu_2)r, \label{NormalVector:BG}
\Eeq
where $\epsilon=\pm1$.
This completely determines $\tau^a=dy^a/d\tau$ where $\tau$ is the
proper time of the curve $\Gamma$ as
\Beqar
&&{dt_\pm\over d\tau}= \epsilon' (n^r)_\pm,\label{DW:EOM1}\\
&&{dr\over d\tau}= \epsilon' (n^t)_\pm,\label{DW:EOM2}
\Eeqar
where $\epsilon'$ is the sign of $n^r$.  In particular from the
continuity of $r$ and the first of Eq.(\ref{NormalVector:BG})
it follows that $\diff{(n^r)^2}=0$, which implies that $\mu_2$ vanishes,
hence
\Beqar
&&\mu_1=\mu_2=0,\\
&& \left(\tau^c D_\para n_c\right)_\pm
= - \left({1\over r}D_\orth r\right)_\pm = \pm \lambda. \label{DWeq}
\Eeqar
This boundary conditions are written in terms of $K^j_k$ as
\Beq
\left(K^j_k\right)_\pm = \pm\lambda \delta^j_k, \label{ReflectionSymmetry}
\Eeq
which implies that the spacetime is reflection symmetric with respect
to the domain wall.
Under this condition Eqs.(\ref{DW:EOM1}) and (\ref{DW:EOM2}) are
easily solved to yield the hyperboloid
\Beq
r^2- t^2 = \lambda^{-2}.
\Eeq

\begin{figure}[t]
\parbox{7cm}{\epsfxsize=7cm \epsfbox{fig1.ps}}\hspace{1cm}
\parbox{5cm}{
\noindent
Fig.1  Conformal Diagram of the Background
Spacetime.\par
\medskip
{\small
\noindent
The entire background spacetime is constructed by identifying the two
boundary hyperboloids $\Sigma_-$ and $\Sigma_+$ of two copies $\M_+$
and $\M_-$ of a region in the Minkowski spacetime.
}}
\end{figure}

Note that Eq.(\ref{DWeq2}) shows that $\N_\pm$ both correspond to the
inside of the hyperboloid. Hence the whole spacetime is spatially
compact and globally hyperbolic, but has null future and past
infinities each of which have two connected components, as shown in
Fig.1.

\section{Perturbation Equations}

In this section, following Gerlach and Sengupta, we introduce gauge-
invariant variables which are local combinations of the tensor
harmonic expansion coefficients of metric perturbation variables, and
express the perturbed Einstein equations in terms of them.  Further we
show that the set of gauge-invariant equations so obtained can be
reduced to a single second-order wave equation in the orbit space $\N$
by introducing a gauge-invariant master variable for the flat
background.  This wave equation is just the flat limit of the Zerilli
equation for the Schwarzschild background\CITE{Regge.T&Wheeler1957,%
Zerilli.F1970,Chandrasekhar.S1983B}.

\subsection{Gauge-invariant variables}

Let $f_{\mu\nu}$ be the perturbation of the metric
\Beq
f_{\mu\nu} := \delta g_{\mu\nu}.
\Eeq
Then under the infinitesimal gauge-transformation $\bar\delta x^\mu = \xi^\mu$,
$f_{\mu\nu}$ transforms as
\Beq
\bar\delta f_{\mu\nu} = -\nabla_\mu \xi_\nu - \nabla_\nu \xi_\mu.
\Eeq
In the coordinate system $(y^a,z^p)$ introduced in the previous section,
with the help of Eqs.(\ref{NablaByD1}) and (\ref{NablaByD2}) this
equation is written as
\Beqar
&&\bar\delta f_{ab} = -D_a\xi_b - D_b\xi_a, \label{GaugeTrf1}\\
&&\bar\delta f_{ap} = -r^2D_a\Frac({\xi_p}/{r^2}) - \hat D_p \xi_a,
\label{GaugeTrf2}\\
&&\bar\delta f_{pq} = -\hat D_p\xi_q -\hat D_q\xi_p- 2r\xi^aD_ar\Omega_{pq}.
\label{GaugeTrf3}
\Eeqar

Let us expand the  perturbation variables $f_{ab}$, $f_{ap}$,
$f_{pq}$, $\xi_a$, and $\xi_p$ by the tensor harmonics(for their
definitions and fundamental properties see Appendix A) as
\Beqar
&& f_{ab} = \sum f_{ab} Y^l_m, \\
&& {1\over r}f_p^a = \sum \left[\evu{f}{a}(V_\even{}^l_m)_p
 + \ovu{f}{a}(V_\odd{}^l_m)_p\right],\\
&& {1\over r^2}f_{pq} = \sum \left[\es{f} (T_\evens{}^l_m)_{pq}
+ \et{f} (T_\event{}^l_m)_{pq} + \ot{f} (T_\oddt{}^l_m)_{pq}\right],\\
&& \xi_a = \sum \xi_a Y^l_m,\\
&& {1\over r}\xi_p = \sum \left[\xi_\even(V_\even{}^l_m)_p
+ \xi_\odd(V_\odd{}^l_m)_p\right].
\Eeqar
In these expressions the summation is over all possible values of
$(l,m)$, and the expansion coefficients are tensors on the orbit
space $\N$.  The indices $l$ and $m$ for the coefficients are omitted
for simplicity. The prefactors $1/r$ or $1/r^2$ on the left hand
sides are put to give the expansion coefficients the same physical
dimension.  The symbol o and e refer to the odd and the even modes,
respectively.

By inspecting the gauge transformation of the coefficients we can
easily find all gauge-invariant combinations.  We can discuss the odd
and the even modes separately.

First for the odd modes Eqs.(\ref{GaugeTrf2}) and (\ref{GaugeTrf3})
yield
\Beqar
&& \bar\delta \ovu{f}{a} = -r D^a\Frac({\xi_\odd}/r) \quad (l\ge1),\\
&& \bar\delta \ot{f} = -{2\over r}\xi_\odd \qquad (l\ge2).
\Eeqar
{}From these equations we immediately see that for $l\ge2$ there exist
two independent gauge-invariant quantities. The most natural choice is
\Beq
F^a := \ovu{f}{a} - {1\over2}r D^a \ot{f}.
\Eeq
On the other hand for $l=1$, since $\ot{f}$ does not exist, there is
only one gauge-invariant which is given by
\Beq
\epsilon^{ab}D_a\left({1\over r}\ovd{f}{b}\right).
\Eeq
This $l=1$ variable expresses  just the total
angular momentum perturbation and has no relevance to gravitational
waves when we impose the Einstein equations.  Though this fact is well
known\CITE{Zerilli.F1970}, We will prove it later for completeness.

Next for the even modes the gauge transformation of the expansion
coefficients are expressed as
\Beqar
&& \bar\delta f_{ab} = - D_a\xi_b - D_b\xi_a, \qquad(l\ge0)\\
&& \bar\delta \evu{f}{a} = -rD^a\Frac({\xi_\even}/r) - {1\over r}\xi^a,
\quad (l\ge1)\\
&& \bar\delta \es{f} = {2l(l+1)\over r}\xi_\even - {4\over r}D^ar\xi_a
\quad (l\ge0)\\
&& \bar\delta \et{f} = -{2\over r}\xi_\even. \qquad (l\ge2)
\Eeqar
{}From these equations the variable $X^a$ defined by
\Beq
X^a := r\evu{f}{a} - {1\over2}r^2D^a\et{f}
\Eeq
transforms as
\Beq
\bar\delta X^a = -\xi^a.  \label{GaugeTrf:X}
\Eeq
Hence for $l\ge2$ we can easily find all the gauge-invariant
combinations.  The most natural independent sets are
\Beqar
&&F_{ab}:= f_{ab} - D_aX_b -D_bX_a,\\
&&F:= \es{f} + l(l+1)\et{f} -{4\over r}D_ar X^a.
\Eeqar

As in the odd modes, the cases $l\le1$ require separate treatments.
The case $l=0$ is simple. From the Birkhoff's theorem, the perturbed
geometry is nothing but a Schwarzschild geometry with a small mass if
the Einstein equations are taken into account. Hence it has no
relevance to gravitational waves.

Though the argument is not so simple, it was shown by Zerilli that
$l=1$ case is also irrelevant\CITE{Zerilli.F1970}.  For completeness
we will prove it later by the gauge-fixing method.

\subsection{Gauge-invariant Perturbation Equations}

In the flat background spacetime the perturbed vacuum Einstein equations
are given by
\Beqar
&2\delta G_{\mu\nu} & \equiv -\Box f_{\mu\nu} -\nabla_\mu\nabla_\nu
+\Box f g_{\mu\nu} + \nabla_\mu\nabla_\lambda f^\lambda_\nu
+\nabla_\nu\nabla_\lambda f^\lambda_\mu
- g_{\mu\nu}\nabla_\lambda\nabla_\sigma f^{\lambda\sigma} \nonumber \\
&& = 0, \label{PEeq:general}
\Eeqar
where $\Box=\nabla_\mu\nabla^\mu$ and $f=f^\mu_\mu$.  Let us rewrite this
set of equations in terms of the gauge-invariant variables introduced just
above and show that they are reduced to a single master equation for each
of the odd and the even modes.

\subsubsection{Odd modes}

First we discuss the exceptional case $l=1$.  For this case
Eq.(\ref{PEeq:general}) reduces to
\Beq
2\ovu{(\delta G)}{a} \equiv D_b\left[r^4D_a\Frac({\ovu{f}{b}}/r)
-r^4D^b\Frac({\ovd{f}{a}}/r)\right]=0.
\Eeq
{}From this it follows that the gauge-invariant is given by
\Beq
\epsilon_{ab}D^a\Frac({\ovu{f}{b}}/r) = {C \over r^4},
\Eeq
where $C$ is a constant.  This is just an angular momentum
perturbation, hence has no relevance to gravitational waves.

Next for the general case $l\ge2$ Eq.(\ref{PEeq:general})
reduces to two sets of equations. First from the odd tensor component we
get
\Beq
r^2 \ot{(\delta G)}\equiv D_a(rF^a) =0. \label{PEeq:o1}
\Eeq
This implies that $F_a$ is expressed in terms of a master variable $\opot$ as
\Beq
rF^a = \epsilon^{ab}D_b\opot.  \label{opot:def}
\Eeq
Second from the odd vector component we get
\Beq
2\ovu{(\delta G)}{a}\equiv -\left(\Box_2-{l(l+1)\over r^2}\right)(rF^a)
+2D^arD_br F^b =0, \label{PEeq:o2}
\Eeq
where $\Box_2:=D_a D^a$.
In terms of  $\opot$ this equation is written as
\Beq
\epsilon^{ab}D_b\left[(r^2\Box_2 -2rD^crD_c + 2-l^2-l)\opot\right]=0.
\Eeq
This implies that the term inside the square bracket is a constant.
For $l\ge2$ this constant can be put to zero by using the freedom of
adding an arbitrary constant in the definition of $\opot$
(\ref{opot:def}). Hence the perturbed Einstein equation is reduced to
the following single equation for $\opot$:
\Beq
\left(-\Box_2 + {l(l+1)\over r^2}\right)
\Frac({\opot}/r)=0. \label{MasterEq:odd}
\Eeq

\subsubsection{Even modes}

We first discuss the generic case $l\ge2$.  In this case
Eq.(\ref{PEeq:general}) yields the following set of gauge-invariant
equations:
\Beqar
&& F^a_a =0, \label{PEeq:e1}\\
&&  D_b F^b_a={1\over2}D_aF, \label{PEeq:e2}\\
&&  \left(\Box_2 + {2\over r}D_arD^a-{l^2+l-2\over r^2}\right)F
= {4\over r^2}D_arD_br F^{ab}, \label{PEeq:e3}\\
&& -\left(\Box_2 + {2\over r}D_crD^c-{l(l+1)\over r^2}\right)F_{ab}
+ {2\over r}D_cr\left(D_aF^c_b + D_b F^c_a\right) \nonumber\\
&& \qquad = {1\over r}(D_ar D_b + D_br D_a)F. \label{PEeq:e4}
\Eeqar
These correspond to the $\et{(\delta G)}$, $\evu{(\delta G)}{a}$, $(\delta
G)^a_a$, and $(\delta G)^a_b-(\delta G)^c_c\delta^a_b/2$ of
Eq.(\ref{PEeq:general}), respectively.  $\es{(\delta G)}$-component does not
yield an independent equation because of the Bianchi identity. This
identity also guarantees the consistency of these equations.

Now we show that the above set of equations can be reduced to the same
single equations as for the odd modes.  First, as shown in Appendix B,
from Eqs.(\ref{PEeq:e1}) and (\ref{PEeq:e2}) we can express $F^a_b$
and $F$ in terms of a scalar master variable $\epot$ as
\Beqar
&& F_{ab} = D_aD_b\epot - {1\over2}\gamma_{ab}\Box_2 \epot, \label{FbyEpot1}\\
&& F = \Box_2 \epot. \label{FbyEpot2}
\Eeqar
Putting these expressions into Eqs.(\ref{PEeq:e3}) and (\ref{PEeq:e4}), we
obtain the following three equations for $\epot$:
\Beq
\partial_u^2 L=0, \quad \partial_v^2 L =0, \quad \partial_u\partial_v L =0,
\Eeq
where
\Beq
L := \left( -r^2\Box_2 + 2r \partial_a rD^a +
l^2+l-2\right)\epot.
\Eeq
These imply that $L$ is a constant.  As in the odd modes this constant can
be put zero by utilizing the freedom of adding arbitrary constant to
$\epot$. Hence the perturbed Einstein equation reduces to the same equation
as that for the odd modes:
\Beq
\left(-\Box_2+ {l(l+1)\over r^2}\right)
\Frac({\epot}/r)=0. \label{MasterEq:even}
\Eeq

Finally we will show that there exists no physical mode for the case $l=1$.
In this case the perturbed Einstein equations are expressed in terms of $F$
and $F_{ab}$ by the same set of equations if we put $\et{f}=0$ in the
definitions of $F$ and $F_{ab}$, except that Eq.(\ref{PEeq:e1})
corresponding to $\et{(\delta G)}$ is missing. However, since $F$ and
$F_{ab}$ are not gauge-invariant any longer and transforms as
\Beqar
&& \bar\delta F_{ab}= 2r D_aD_b\xi_\even, \quad (l=1),\label{GaugeTrf1(l=1)}\\
&& \bar\delta F = 4 D^a r D_a\xi_\even, \quad(l=1) \label{GaugeTrf2(l=1)}
\Eeqar
we can supplement Eq.(\ref{PEeq:e1}) as a gauge
condition. Under this gauge condition the same argument as in the generic
case $l\ge2$ can be applied.  What is different in the present case is that
$\epot$ is not gauge invariant because the condition $F^a_a=0$ does not
completely fix the gauge.  The  residual gauge freedom is expressed as
\Beq
\xi_\even = f(u) + g(v),
\Eeq
where $f(u)$ and $g(v)$ are arbitrary functions of $u$ and $v$,
respectively. $\epot$ transforms under this gauge transformation as
\Beq
\bar\delta \epot = 2r^2\partial_r\left[ {1\over r} \left(-\int du f(u) +
\int dv g(v) \right)\right].
\Eeq
However, it is easily checked that the right hand side of this equation is
the general solution of Eq.(\ref{MasterEq:even}) for $l=1$.  This implies
that all the gauge invariants vanishes for $l=1$ solutions to the perturbed
Einstein equations.

\subsection{Solutions to the master equation}

As we have seen in the previous subsection, the perturbed Einstein
equations for the odd and the even modes reduce to the same wave equation,
Eq.(\ref{MasterEq:odd}) or Eq.(\ref{MasterEq:even}).  In this subsection we
construct the general solution to it by  the method of separation of
variables in the Minkowski coordinates and in the Rindler coordinates,
and give the relation between the mode functions in both coordinates.
Throughout this subsection we denote the
master  variable simply by $\Phi$.

First in the Minkowski coordinates, the master equation is simply given by
\Beq
\left(\partial_t^2-\partial_r^2+{l(l+1)\over r^2} \right)\Frac(\Phi/r)=0.
\Eeq
With the help of the Fourier transform with respect to $t$, the general
solution to this equation that is regular at $r=0$ and belongs to
$L^2(-\infty,\infty)$ with respect to $t$ for any fixed value of $r$
is written as
\Beq
\Phi = r^2 \int_{-\infty}^\infty d\omega\, C(\omega) e^{-i\omega t}
j_l(\omega r),
\label{ModeExpansion:Minkowski}
\Eeq
where $j_l(x)$ is the spherical Bessel function. From the orthonormality
relation
\Beq
2\omega\omega'\int_0^\infty dr\,r^2j_l(\omega r)j_l(\omega' r)
=\delta(\omega-\omega') + (-1)^{l+1}\delta(\omega+\omega'),
\label{Orthonormality:Bessel}
\Eeq
the mode expansion coefficient $C(\omega)$ is written in terms of the
initial data of $\Phi$ at $t=0$ as
\Beq
C(\omega)=\omega \int_0^\infty dr j_l(\omega r)
\left.(\omega\Phi+i\partial_t\Phi)\right|_{t=0}.
\label{CbyPhi}
\Eeq

Next in the Rindler coordinates,
\Beq
t=\chi\sinh\tau, \qquad r=\chi\cosh\tau,
\Eeq
which cover the region $|t|<r$, the master equation is written as
\Beq
\left(-\chi^2\partial_\chi^2 -\chi\partial_\chi +\partial_\tau^2
+{l(l+1)\over\cosh^2\tau}\right)\Frac(\Phi/r) =0.
\Eeq
If $\Phi/r$ is written as $R(\chi)T(\tau)$, $R(\chi)$ and $T(\tau)$ satisfy
the equations
\Beqar
&&\left(\partial_\chi^2+{1\over\chi}\partial_\chi + {k^2\over\chi^2}\right)
R(\chi) =0,\\
&&\left(\partial_\tau^2 + {l(l+1)\over\cosh^2\tau}+ k^2\right)T(\tau)=0,
\label{EqForT}
\Eeqar
where $k$ is a separation constant. The solutions to these equations are
given by
\Beqar
&&R(\chi) \propto \chi^{\pm ik} = e^{\pm ik \ln\chi}, \\
&&T(\tau) \propto P^{\pm ik}_l(\tanh\tau)
\propto G_l\left(1,1\mp ik;{1+\tanh\tau\over 2}\right)e^{\mp ik\tau},
\Eeqar
where $P_l^\mu(x)$ and $G_l(a,b;x)$ are the Legendre function and the
Jacobi polynomial.

{}From Eq.(\ref{EqForT}) we can easily show that the set of functions
$\{\psi_l(\tau,k)\}$ defined by
\Beq
\psi_l(\tau,k) := G_l\left(1,1- ik;{1+\tanh\tau\over 2}\right)e^{- ik\tau}
\label{NormalMode:Rindler}\Eeq
satisfies the orthogonality condition
\Beq
\int^\infty_{-\infty} \overline{\psi_l(\tau,k')}\psi_l(\tau,k)d\tau = 2\pi
\delta(k-k').
\Eeq
Further from the general theorem it can be shown that $\{\psi_l(\tau,k)\}$
($-\infty<k<\infty$) is complete in the $L^2(-\infty,\infty)$ space of
functions of $\tau$.  Hence if $\Phi/r$ belongs to $L^2(-\infty,\infty)$ for
any fixed value of $\chi(>0)$, it is expressed as
\Beqar
&\displaystyle{\Phi\over r} & = \int^\infty_{-\infty} dk
\psi_l(\tau,k)\left(A(k)e^{ik\ln\chi}+B(k)e^{-ik\ln\chi}\right) \nonumber\\
&& = \int^\infty_{-\infty} dk
G_l\left(1,1-ik;{1+\tanh\tau\over2}\right) \left(A(k)e^{-ik\ln(-u)}
+B(k)e^{-ik\ln v}\right). \nonumber\\
&& \label{ModeExpansion:Rindler}
\Eeqar

In connecting the solutions in the two regions $\M_+$ and $\M_-$ along the
domain wall, the expression (\ref{ModeExpansion:Rindler}) is more useful.
However, it is not appropriate for studying the behavior of the solution
in the whole spacetime region since it is restricted in the Rindler wedge
$|t|<r$.  Obviously the expression (\ref{ModeExpansion:Minkowski}) should
be used for that purpose.  Hence we need the relation between the coefficients
in these two expansions in order to construct the global solution and study
its behavior.  This relation is easily obtained by substituting
(\ref{ModeExpansion:Rindler}) into Eq.(\ref{CbyPhi}):
\Beqar
&C(\pm|\omega|) & = e^{-il\pi/2}\int_{-\infty}^\infty dk\, e^{\pm k\pi/2}
\Gamma(1-ik) \nonumber\\
&&\quad \times \left(A(k)|\omega|^{-ik}{\Gamma(l+1+ik)\over\Gamma(l+1-ik)}
+B(k)|\omega|^{ik}\right).  \label{CbyAB}
\Eeqar

\begin{figure}[t]
\parbox{4cm}{\epsfxsize=4cm \epsfbox{fig2.ps}}\hspace{2cm}
\parbox{7cm}{
\noindent
Fig.2  Relation between the Rindler modes and the
Mikowskian modes.
}

\end{figure}

The meaning of this relation becomes clearer if we rewrite it as the
relation between the asymptotic value of $\Phi$ at the future null
infinity $\I^+$ and the past null infinity $\I^-$. First note that
$\Phi/r$ has a finite limit at null infinities:
\Beq
{\Phi\over r} \rightarrow \int^\infty_{-\infty} {d\omega\over2\omega}
\left(e^{-i\omega u-i{l+1\over2}\pi}+e^{-i\omega v+i{l+1\over2}\pi}\right)
C(\omega).
\Eeq
Hence if we define the asymptotic values of $\Phi/r$ at $\I^-$ and $\I^+$
by
\Beq
\alpha(v) := \left.(\Phi/r)\right|_{\I^-}, \qquad
\beta(u) := \left.(\Phi/r)\right|_{\I^+},
\Eeq
they are expressed in terms of $C(\omega)$ as
\Beqar
&&\alpha(v) = {1\over2}e^{i{l+1\over2}\pi}\int^\infty_{-\infty}
{d\omega\over\omega}C(\omega)e^{-i\omega v}, \\
&&\beta(u) = {1\over2}e^{-i{l+1\over2}\pi}\int^\infty_{-\infty}
{d\omega\over\omega}C(\omega)e^{-i\omega u}.
\Eeqar
By inserting the expression (\ref{CbyAB}) into the right hand sides of
these equations, we get after a short calculation
\Beqar
&&\alpha(v)=\theta(-v)\pi \int^\infty_{-\infty}dk\,A(k)|v|^{ik}
{\Gamma(-ik)\over\Gamma(ik)}{\Gamma(l+1+ik)\over\Gamma(l+1-ik)}\nonumber\\
&&\quad +\theta(v)\pi \int^\infty_{-\infty}dk\,B(k)v^{-ik},
\label{alphaByAB}\\
&&\beta(u)=\theta(-u)(-1)^{l+1}\pi \int^\infty_{-\infty}dk\,A(k)|u|^{ik}
{\Gamma(-ik)\over\Gamma(ik)}{\Gamma(l+1+ik)\over\Gamma(l+1-ik)}\nonumber\\
&&\quad +\theta(u)\pi(-1)^{l+1}\int^\infty_{-\infty}dk\,B(k)u^{-ik}.
\label{betaByAB}
\Eeqar
These equations show that $A(k)$ and $B(k)$ describe the freedoms of
the incidental gravitational waves from the part$(v<0)$ of $\I^-$ outside the
Rindler wedge and the outgoing ones to the part$(u>0)$ of $\I^+$ outside the
Rindler wedge, respectively.  This situation is schematically depicted
in Fig.2.

\section{Global Solutions}

In this section we construct global solutions to the perturbed Einstein
equations in $\M$ by matching the solutions in $\M_+$ and $\M_-$ along the
domain wall, and study the behavior of gravitational waves on the whole
spacetime.

Like the background geometry the global solutions to the perturbed
Einstein equations must satisfy the perturbed junction condition
\Beqar
&&\diff{\delta q_{jk}}=0, \label{PJC1}\\
&&\diff{\delta K^j_k}=\kappa^2 \left(\delta S^j_k - {1\over2}\delta^j_k
\delta S^l_l\right). \label{PJC2}
\Eeqar
We get the boundary conditions for the global solutions to satisfy, if we
express these conditions in terms of the gauge-invariant variables and
rewrite them as conditions on the boundary values of the master variables.
The most cumbersome part of this procedure is to express $\delta K^j_k$
in terms of the gauge-invariant variables.  This was already done by
Gerlach and Sengupta\CITE{Gerlach.U&Sengupta1979a,%
Gerlach.U&Sengupta1979b}.  Their result specialized
to our case is written in our notations as:
\Beqar
& \ovu{(\delta K_\para)}{a}\tau_a &= {1\over2}\epsilon^{ab}
D_b\Frac({F_a}/r), \label{deltaK:odd1}\\
& \ot{(\delta K_\para)} &= {1\over r} F_\orth, \label{deltaK:odd2}\\
& (\delta K_\para)^a_a &= {1\over2}K^a_a F_{\para\para}
- {1\over2}D_\para F_{\para\orth}
 -{1\over2}D_\orth F^a_a + {1\over2}n_a D_b F^{ab}  \nonumber\\
&& \quad + \{ -D_\para^2 + (K^a_a)^2\} X_\orth, \label{deltaK:even1}\\
& \evu{(\delta K_\para)}{a}\tau_a & = {1\over 2r}F_{\para\orth}
+ D_\para\Frac({X_\orth}/r), \label{deltaK:even2}\\
& \es{(\delta K_\para)} &= -{1\over2}K^p_p F_{\orth\orth}
- {2\over r}D_\para r F_{\para\orth} - {1\over 2}D_\orth F \nonumber\\
&& \quad + \left\{-{2\over r}D_\para r D_\para - {l(l+1)\over r^2}
+ {1\over2}(K^p_p)^2\right\}X_\orth, \label{deltaK:even3}\\
& \et{(\delta K_\para)} & = {1\over r^2}X_\orth, \label{deltaK:even4}
\Eeqar
where $(\delta K_\para)$ denotes the four-dimensional extension of the
tensor $\delta K^j_k$ given by
\Beq
(\delta K_\para)^\mu_\nu := q^\mu_\lambda \delta K^\lambda_\nu,
\Eeq
the symbols $V_\para$ and $V_\orth$ for a vector $V_a$ on $\N$ denote
\Beq
V_\para := \tau^a V_a, \quad V_\orth:= n^a V_a,
\Eeq
and $T_{\para\para}$, $T_{\para\orth}$ and $T_{\orth\orth}$ for a symmetric
tensor $T_{ab}$ on $\N$ denote
\Beq
T_{\para\para}:=\tau^a\tau^b T_{ab}, \quad
T_{\para\orth}:=\tau^a n^b T_{ab}, \quad
T_{\orth\orth}:=n^an^b T_{ab}.
\Eeq

The reader may notice that there appears the variable $X^a$ on the
above expression for $\delta K^j_k$ which was treated as
gauge-dependent variable in the previous section.  We should comment
on this point.  In the perturbation theory we pull back the perturbed
geometry to the unperturbed background geometry by some embedding. The
ambiguity in the choice of this embedding is the gauge freedom.  When
one considers only regular spacetimes, there is no a priori
restriction on the gauge choice apart from the topological one.
However, when one considers spacetimes with singularities such as the
domain wall, the perturbative treatment obviously breaks down unless
the embedding preserves the singularity structure.  Specified to our
problem, this implies that we must restrict the embeddings so that it
maps the domain wall in the background spacetimes to that in perturbed
spacetimes.  In terms of the gauge transformation this restriction is
expressed as $\xi_\orth=\xi_a n^a=0$ on the domain wall.  Hence from
Eq.(\ref{GaugeTrf:X}) $X_\orth$ becomes gauge invariant on the domain
wall, and in a sense describes the perturbation of the location of the
domain wall.

This treatment of the singular surface is slightly different from that
by Gerlach and Sengupta.  Instead of the physical gauge fixing, they
introduced a scalar field $\phi$ defined on the whole spacetime,
and described the singular surface as a hypersurface on which $\phi$ takes
a constant value.  In this approach
$\Delta\phi:=\delta \phi - D_\orth\phi X_\orth$, where
$\orth$ is defined by the unit normal to $\phi$=const surface, is gauge
invariant and its value on the singular surface describes the perturbation
of its location.  Though this approach is essentially equivalent to ours
because $\Delta\phi/D_\orth\phi = -X_\orth$ on the domain wall in the
above physical gauge, our treatment seems to be more economical in that
we do not need to introduce an extra variable with a redundant freedom.

In order to fix the junction condition completely,  we must specify the
perturbation of the surface energy-momentum density $S^j_k$ of the domain
wall. In this paper we assume that $S^j_k$ keeps the form
(\ref{EnergyMomentum:DW}) under perturbation. This generally leads to
\Beq
\delta S^j_k = -4 \delta\lambda \delta^j_k. \label{deltaS:DW}
\Eeq
The unknown $\delta\lambda$ is strongly restricted by the consistency of
the perturbed junction condition. In fact, from  Eqs.(\ref{PJC1}) and
(\ref{PJC2}), and the perturbation of the momentum constraint
\Beq
\diff{\threeD\nabla_k \delta K^k_j}=0,
\Eeq
it follows that $\threeD\nabla_j \delta\lambda=0$.  This implies that $\delta
S^j_k$ affects only on the $l=0$ mode which is just the mass perturbation
as explained in the previous section.  Thus we can simply set
\Beq
\delta S^j_k=0,
\Eeq
for which the second junction condition (\ref{PJC2}) is written as
\Beq
\diff{\delta K^j_k}=0.\label{PJC2'}
\Eeq

Now let us rewrite the above junction condition in terms of the master
 variable and construct the global solutions.  We discuss
the odd and the even modes separately.

\subsection{Odd modes}

First from Eq.(\ref{PJC1}) $\ot{f}$ and $\ovd{f}{\para}$ is continuous
across the domain wall.  Hence we get
\Beq
\diff{F_\para} =\diff{\ovd{f}{\para}-{r\over2}D_\para\ot{f}}=0.
\Eeq
Since $F^a$ is written from Eq.(\ref{opot:def}) as
\Beq
rF^a = -\tau^a D_\orth\opot + n^aD_\para \opot,
\Eeq
this is expressed in terms of $\opot$ as
\Beq
\diff{D_\orth\opot}=0. \label{PJC:opot1}
\Eeq
Next from Eqs.(\ref{deltaK:odd2}) and (\ref{PJC2'}) we get
\Beq
\diff{\epsilon^{ab}D_b\left({1\over r}F_a\right)}
={1\over r^2}\diff{\left(\Box_2-{2\over r}D^arD_a\right)\opot}=0.
\Eeq
Hence from the master equation (\ref{MasterEq:odd}) we obtain
\Beq
\diff{\opot}=0. \label{PJC:opot2}
\Eeq
No new constraint is obtained from the other components of
Eq.(\ref{PJC2'}).

Since Eq.({\ref{PJC:opot1}) is expressed from Eq.(\ref{DWeq}) as
\Beq
\mean{\partial_\chi\Frac(\opot/r)}=-\lambda {\mean{\opot}\over r},
\label{PJC:opot1'}
\Eeq
the junction conditions on the solution are written in terms of the mode
expansion coefficients $A$ and $B$ in Eq.(\ref{ModeExpansion:Rindler})
as
\Beqar
&& \diff{A} + \diff{B} = 0,\\
&& (k-i)\mean{A} - (k+i)\mean{B}=0.
\Eeqar
These equations are easily solved to yield
\Beq
B_\pm = - {i\over k+i} A_\pm + {k\over k+i} A_\mp.
\Eeq

We can express this result as a relation of the asymptotic values of
$\opot/r$ at the null infinities of the total spacetime, $\alpha_\pm(v)$
($v<0$)
and $\beta_\pm(u)$($u>0$), with the help of Eqs.(\ref{alphaByAB}) and
(\ref{betaByAB}):
\Beqar
&\beta_\pm(u) &= (-1)^{l+1}\Big\{(-1)^l\alpha_\mp(-1/u) \nonumber\\
&& \quad + \sum_{n=1}^l{n\over n+1}{{}_{l+n}C_l\over (n-1)!}u^n
\int^0_{-1/u} dv\,v^{n-1}\alpha_\mp(v) \nonumber\\
&& \quad - \sum_{n=1}^l{1\over n+1}{{}_{l+n}C_l\over (n-1)!}u^n
\int^0_{-1/u} dv\,v^{n-1}\alpha_\pm(v)\Big\}. \label{alphaBybeta:odd}
\Eeqar
{}From this we find that $\beta_\pm(u)$ depends only on the values of
$\alpha(v)$ at $v\ge-1/u$.  Since $-1/v$ is the maximum value  of $u$
on the light cone emitted at $v$ of $\I^-$, this implies that the domain
wall emits gravitational waves only during the period when the incoming
waves hit it, hence that the domain wall has no dynamical degree of freedom.
This result is natural because there exists no odd variable describing
the position of the domain wall.

\subsection{Even Modes}

First from Eq.(\ref{PJC1}) $f_{\para\para}$, $\evd{f}{\para}$, $\es{f}$ and
$\et{f}$ are continuous across the domain wall $\Sigma$.  From this it
follows that
\Beq
\diff{X_\para}=\diff{r\evd{f}{\para} - {1\over2}r^2D_\para \et{f}}=0.
\label{PJC:even1}
\Eeq
On the other hand from Eq.(\ref{deltaK:even4}) and Eq.(\ref{PJC2}) $X_\orth$
is also continuous across $\Sigma$:
\Beq
\diff{X_\orth}= r^2\diff{\et{(\delta K_\para)}} =0. \label{PJC:even2}
\Eeq

Noting this continuity of $X_a$ , we can calculate the jump of values of
$F$ and $F_{\para\para}$ as
\Beqar
&& \diff{F} = \diff{\es{f}+l(l+1)\et{f} - {4\over r}D_a rX^a}
= 8\lambda X_\orth, \label{PJC:even3} \\
&& \diff{F_{\para\para}}=\diff{f_{\para\para}-2D_\para X_\para-2K^a_a X_\orth}
=-4\lambda X_\orth. \label{PJC:even4}
\Eeqar
Further from Eqs.(\ref{deltaK:even2}) and (\ref{deltaK:even3}) the condition
(\ref{PJC2'}) for the components $(e1)$ and $(e0)$ yields
\Beqar
&& \diff{F_{\orth\para}}=0, \label{PJC:even5} \\
&& \diff{D_\orth F}= -4\lambda \mean{F_{\orth\orth}}. \label{PJC:even6}
\Eeqar
Using Eqs.(\ref{PEeq:e1}) and (\ref{PEeq:e2}), we see that
no new constraint is obtained from the continuity of $(\delta K_\para)^a_a$.

Now we rewrite these junction conditions
Eqs.(\ref{PJC:even3})-(\ref{PJC:even6}) as conditions on the master
 variable $\epot$. First from Eq.(\ref{FbyEpot2}) and the master equation
 (\ref{MasterEq:even}), $\diff{F}$ is expressed as
\Beq
\diff{F}=2\cosh\tau\diff{\partial_\chi\Frac(\epot/r)}
-\lambda\left(2\sinh\tau\partial_\tau-{l(l+1)\over\cosh\tau}\right)
\diff{\epot\over r}.  \label{diffF}
\Eeq
Similarly from Eq.(\ref{FbyEpot1}), $\diff{F_{\para\para}}$ is expressed as
\Beq
\diff{F_{\para\para}}=\lambda\left(\cosh\tau\partial_\tau^2
+\sinh\tau\partial_\tau+{l(l+1)\over 2\cosh\tau}\right)\diff{\epot\over r}.
\label{diffFpara}
\Eeq
{}From Eq.(\ref{PJC:even3}) and Eq.(\ref{diffF}) we find that $X_\orth$ is
completely determined by the jump of $\epot$ and its $\chi$-derivative
at $\Sigma$ as
\Beq
X_\orth = {1\over 4\lambda}\cosh\tau\diff{\partial_\chi\Frac(\epot/r)}
-{1\over4}\left(\sinh\tau\partial_\tau-{l(l+1)\over2\cosh\tau}\right)
\diff{\epot\over r}.  \label{XorthByEpot}
\Eeq
Further elimination of $X_\orth$ from Eqs.(\ref{PJC:even3}) and
(\ref{PJC:even4}) yields the boundary condition on $\epot$,
\Beq
\left(\partial_\tau^2 + {l(l+1)\over\cosh^2\tau}\right)\diff{\epot\over r}
=-{1\over\lambda}\diff{\partial_\chi\Frac(\epot/r)}. \label{PJC:epot1}
\Eeq

Next from Eq.(\ref{FbyEpot1}) $\diff{F_{\para\orth}}$ and
$\mean{F_{\orth\orth}}$ are expressed as
\Beqar
&&\diff{F_{\para\orth}}= -2\partial_\tau\left\{\cosh\tau
\mean{\partial_\chi\Frac(\epot/r)}\right\} \label{ForthByEpot}\\
&&\mean{F_{\orth\orth}}=\lambda\left(\cosh\tau\partial_\tau^2
+\sinh\tau\partial_\tau+{l(l+1)\over2\cosh\tau}\right)
\mean{\Frac(\epot/r)}.  \label{FparaByEpot}
\Eeqar
Similarly from Eq.(\ref{FbyEpot2}) and Eq.(\ref{MasterEq:even}),
the jump of $D_\orth F$ is expressed as
\Beqar
&\diff{D_\orth F} & = 4\lambda\tanh\tau\partial_\tau
\left\{\cosh\tau\mean{\partial_\chi\Frac(\epot/r)}\right\}
-2\lambda{l^2+l-2\over\cosh\tau}\mean{\partial_\chi\Frac(\epot/r)} \nonumber\\
&& \quad -4\lambda^2\left(\cosh\tau\partial_\tau^2+\sinh\tau\partial_\tau
+{l(l+1)\over2\cosh\tau}\right)\mean{\Frac(\epot/r)}. \label{DFbyEpot}
\Eeqar

{}From Eq.(\ref{ForthByEpot}), condition (\ref{PJC:even5}) is written as
\Beq
\partial_\tau\left\{\cosh\tau\mean{\partial_\chi\Frac(\epot/r)}\right\}
=0.
\Eeq
On the other hand from Eqs.(\ref{FparaByEpot}) and (\ref{DFbyEpot}),
the last condition (\ref{PJC:even6}) is written as
\Beq
2\sinh\tau\partial_\tau
\left\{\cosh\tau\mean{\partial_\chi\Frac(\epot/r)}\right\}
= -(l^2+l-2)\mean{\partial_\chi\Frac(\epot/r)}.
\Eeq
For $l\ge2$ these two equations reduce to the single equation
\Beq
\mean{\partial_\chi\Frac(\epot/r)}=0. \label{PJC:epot2}
\Eeq

Equations (\ref{PJC:epot1}) and (\ref{PJC:epot2}) yield the boundary
conditions for $\epot$ to satisfy on the domain wall.  As in the case
of odd modes, these conditions can be easily translated into conditions
on the mode expansion coefficient as
\Beqar
&&k\left((k-i)\diff{A}+(k+i)\diff{B}\right)=0, \label{PJC:AB:even1}\\
&&k(\mean{A}-\mean{B})=0. \label{PJC:AB:even2}
\Eeqar
Hence for $k\not=0$ $B_\pm$ is expressed in terms of $A_\pm$ as
\Beq
B_\pm={i\over k+i}A_\pm + {k\over k+i}A_\mp.
\Eeq

As in the odd modes, we can rewrite this equation as a relation
between the amplitudes of the incoming and the outgoing waves. The result
is given by
\Beqar
&\beta_\pm(u) &= (-1)^{l+1}\Big\{(-1)^l\alpha_\mp(-1/u) \nonumber\\
&& \quad - \sum_{n=1}^l{n\over n+1}{{}_{l+n}C_l\over (n-1)!}u^n
\int^0_{-1/u} dv\,v^{n-1}\alpha_\mp(v) \nonumber\\
&& \quad - \sum_{n=1}^l{1\over n+1}{{}_{l+n}C_l\over (n-1)!}u^n
\int^0_{-1/u} dv\,v^{n-1}\alpha_\pm(v)\Big\}. \label{alphaBybeta:even}
\Eeqar
Since this equation is the same as Eq.(\ref{alphaBybeta:odd}) apart
from the difference of the sign of the second term, we can conclude
that the domain wall has no dynamical degree of freedom for the even
perturbations either. Note that this does not mean that the wall
behaves as a rigid surface.  In fact we can easily see that the
variable $X_\orth$, which can be interpreted as describing the
displacement of the wall, take non-zero values while incidental waves
go across the wall. Hence, in contrast to the odd-mode case, this
result is non-trivial.

\section{Discussion}

In this paper we have analyzed the behavior of gravitational waves and
their interaction with a domain wall by the first-order perturbation
theory on a special background solution.  The main conclusion is that
the domain wall does not emit gravitational waves spontaneously by its
free oscillations. This  conclusion is quite embarrassing
because in the treatment neglecting gravity the domain wall has its
own dynamical freedom and its oscillation is expected to become a
source of gravitational waves even in the first order with respect to
its oscillation amplitude.

Of course it is possible that this peculiar result is due to the
special symmetries of our background geometry.  For example, the
spherical symmetry may make the spacetime solutions describing the
emission of gravitational waves by the domain wall singular at origin
because of focusing and blueshift due to the motion of the domain
wall.  Actually we can find two independent even-mode solutions for
each $l$ and $m$ which satisfy the field equations as well as the
junction conditions, but are singular at $r=0$.  They are given by
$\epot=C rP_l(t/r) + D P'_l(t/r)$ for $t>-r$ with arbitrary constants
$C$ and $D$, and $\epot=0$ for $t<-r$ which represents the no-incoming
wave condition.  They correspond to the normal modes
(\ref{NormalMode:Rindler}) with $k=0$ and $k=-i$.  From
Eq.(\ref{PJC:AB:even1}) it is easy to see that these solutions satisfy
the junction condition.  Since the singularity at $r=0$ can be
interpreted to be caused by focusing and the discontinuity at $r=-t$
by the infinite blue shift with respect to the Minkowskian frame due
to asymptotic null behavior of the domain wall, they can be candidates
representing the spontaneous emission of gravitational waves by the
domain wall.

This simple interpretation, however, does not seem to be correct for
the following reasons.  Firstly for the singular solution
corresponding to $k=0$ the junction condition does not restrict the
relation between $C_+$ and $C_-$, hence we can construct a solution
for which the gravitational waves are emitted toward only one of the
two Minkowskian regions.  This is quite unphysical.  Further this
solution also satisfies the junction condition for the odd mode.
Since the motion of the domain wall is described by a displacement
perpendicular to the wall in the flat limit which is an even quantity,
it is expected that the odd mode is decoupled from the motion of the
wall in the first order.  Thirdly, as shown in section 4, the domain
wall immediately stops oscillation after the incidental gravitational
wave packet goes through the wall.  This implies that the wall behaves
as if it has no inertia, hence that it has no intrinsic dynamical
freedom.

Here note that the last result comes from a non-trivial
interaction between the domain wall and gravitational waves through
the junction condition.  In fact, from Eqs.(\ref{ReflectionSymmetry})
and (\ref{PJC2'}), the perturbation of the
junction condition (\ref{diffR}) yields the condition
\Beq
\delta\mean{K}=0.
\Eeq
{}From Eq.(\ref{deltaK:even1}), Eq.(\ref{deltaK:even3}) and the field
equations (\ref{PEeq:e1}) and (\ref{PEeq:e2}), this condition
is written as
\Beq
\left(-D_\para^2-{2\over r}D_\para r D_\para -{l(l+1)\over
r^2}+4\lambda^2\right)X_\orth =
{1\over2r^4}D_\para(r^4\mean{F_{\para\orth}}) +{1\over4}\mean{D_\orth
F}. \label{EOM:DW:curved}
\Eeq
This equation should be compared with the perturbative equation of
motion of a domain
wall in a flat background when the gravity of the wall is neglected.
As shown by Garriga and Vilenkin\CITE{Garriga.J&Vilenkin1991}, the
perpendicular displacement $\phi$ from a stationary configuration of
the wall satisfies a Klein-Gordon type wave equation on the wall,
\Beq
\left(\Box_3 + \epsilon^2\right)\phi=0, \label{EOM:DW:flat}
\Eeq
where $\Box_3$ is the d'Alembertian on the wall and $\epsilon$ is the
difference of the vacuum energies on the two sides of the wall.
Since the left hand side of Eq.(\ref{EOM:DW:curved}) coincides with
the harmonic expansion of d'Alembertian on the hyperboloid apart from
the negative mass term, these two
equations have quite similar structures except for the gravitational
force term in our case. In particular Eq.(\ref{EOM:DW:curved})
apparently shows that oscillations of the wall can propagate along it
freely except for possible growth or decay due to the effective
negative mass term if the gravitational field is fixed. However, this
apparent dynamics is lost if we take all the constraints and the field
equations into account because $X_\orth$ is completely determined by
the boundary values of the master variable $\epot$ representing the
gravitational wave freedom through Eq.(\ref{XorthByEpot}).
It is easy to  show with the help of
Eqs.(\ref{ForthByEpot}), (\ref{FparaByEpot}), (\ref{FbyEpot1}) and
(\ref{FbyEpot2}) that the expression (\ref{XorthByEpot}) for $X_\orth$
automatically satisfies (\ref{EOM:DW:curved}) if the boundary
condition (\ref{PJC:epot1}) is satisfied. Thus the dynamical freedom
of the wall is lost due to the consistency between the gravitational
perturbations of geometry and the motion of the domain wall.

Besides the spherical symmetry, our background solution has some
special  features such as the planar symmetry and the thin wall
nature.  However, we cannot think of a physical mechanics by
which these features prohibit the spontaneous emission of
gravitational waves. Further we can show that for more general types of
perturbations to  $S^j_k$ than Eq.(\ref{deltaS:DW}) the wall can emit
gravitational waves spontaneously.  Hence there is a possibility that
the same result holds for a more general class of background
spacetimes if the perturbation of $S^j_k$ is given by
Eq.(\ref{deltaS:DW}).  To see whether this expectation is true or not,
the case of the spherically symmetric background with a
non-planar-symmetric domain wall is under investigation.

\section*{Acknowledgements}

The authors would like to thank A. Vilenkin for valuable discussion.
This work was supported by the Grant-in-Aid for Science Research Fund
of the Ministry of Education, Science and Culture No.05640340(H.K.).

\appendix

\section{Tensor Harmonics}

The tensor harmonics are orthogonal complete sets of tensors on the
unit Euclidean sphere which are solutions to the eigenvalue problem
 of the Laplace-Beltrami operator.  In this appendix we give the
definitions and the basic properties of the tensor harmonics used
in the text.  Throughout this appendix $\theta$ and $\phi$ denote
the standard angular coordinates of the unit sphere with the metric
\Beq
ds^2 = q_{jk}dx^jdx^k=d\theta^2 + \sin^2\theta d\phi^2,
\Eeq
$D_j$ the covariant derivative with respect to this metric and
$\delta_2 =D_jD^j$.

In the argument of gravitational waves the bases of the tensor harmonics
are often chosen so that each has a definite parity with respect to
the transformation $R:(\theta,\phi) \rightarrow (\pi-\theta,\phi+\pi)$ which
maps each point on the unit sphere to its antipodal point.  Under this
choice the tensor harmonics $T$ which transform as
\Beq
R^*T=(-1)^lT
\Eeq
are called even, and those which transform as
\Beq
R^*T=(-1)^{(l+1)}T
\Eeq
are called odd.

\subsection{Scalar Harmonics}

The scalar harmonics are the spherical harmonic functions $Y^m_l(\theta,\phi)$
satisfying the equations
\Beqar
&&[ \Delta_2 + l(l+1)]Y^m_l=0,\\
&& \partial_\phi Y^m_l = im Y^m_l.
\Eeqar
To be explicit, they are expressed in terms of the Legendre functions as
\Beq
Y^m_l(\theta,\phi)=\sqrt{(2l+1)(l-m)!\over 4\pi(l+m)!}
P^m_l(\cos\theta)e^{im\phi},
\Eeq
and transform under $R$ as
\Beq
(R^*Y^m_l)(\theta,\phi)=Y^m_l(\pi-\theta,\phi+\pi)=(-1)^lY^m_l(\theta,\phi).
\Eeq
Hence these are all even.

\subsection{Vector Harmonics}

On the unit sphere any vector field $v^j$ are written in terms of two scalar
functions $v$ and $w$ as
\Beq
v^j = D^jv + \epsilon^{jk}\partial_k w,
\Eeq
where $\epsilon^{jk}$ is the two-dimensional Levi-Civita tensor.  Since
$\epsilon^{jk}$ is an odd quantity, this implies that any vector can be
expanded by the even vector harmonics $V_\even{}^m_l$ defined by
\Beq
(V_\even{}^m_l)_j := D_j Y^m_l,
\Eeq
and the odd vector harmonics $V_\odd{}^m_l$ defined by
\Beq
(V_\odd{}^m_l)_j  := \epsilon_{jk}D^k Y^m_l.
\Eeq
{}From these definitions it immediately follows that they satisfy
\Beqar
&&(\Delta_2 + l^2+l-1)(V^m_l)^j =0 \quad (\r{even\ and\ odd}), \\
&& D_j(V_\even{}^m_l)^j = -l(l+1) Y^m_l,\\
&& D_j(V_\odd{}^m_l)^j=0,\\
&& D_{[j}(V_\even{}^m_l)_{k]}=0,\\
&& D_{[j}(V_\odd{}^m_l)_{k]}={1\over2}l(l+1)\epsilon_{jk}Y^m_l.
\Eeqar

Note that the vector harmonics vanish for $l=0$,  hence they have
meaning only for $l\ge1$.

\subsection{2nd-rank Tensor Harmonics}

Any smooth symmetric 2nd-rank tensor field $t^{jk}$ on the unit sphere
can be expressed in  terms of its trace $t=t^j_j$ and two scalar fields
$v$ and $w$ as
\Beq
t^{jk}={1\over2}tq^{jk}+ D^jD^k v-{1\over2}q^{jk}\Delta_2 v
+{1\over2}\left(\epsilon^{kn}D^jD_n w + \epsilon^{jn}D^kD_n w\right).
\Eeq
Hence $t^{jk}$ is expanded in terms of the two even tensor harmonics
$\es{T}{}^m_l$ and $\et{T}{}^m_l$ defined by
\Beqar
&&(\es{T}{}^m_l)_{jk} := {1\over2}Y^m_l q_{jk},\\
&&(\et{T}{}^m_l)_{jk} := \left(D_jD_k + {1\over2}l(l+1)q_{jk}\right)Y^m_l,
\Eeqar
and the odd tensor harmonics $\ot{T}{}^m_l$ defined by
\Beq
(\ot{T}{}^m_l)_{jk} :={1\over2}(\epsilon_{kn}D_j+\epsilon_{jn}D_k)D_nY^m_l.
\Eeq

$\es{T}{}^m_l$ is essentially of the scalar type and satisfies
\Beqar
&&[\Delta_2 + l(l+1)](\es{T}{}^m_l)_{jk}=0,\\
&&(\es{T}{}^m_l)^j_j=Y^m_l,\\
&&D_k(\es{T}{}^m_l)^{jk}={1\over2}(V_\even{}^m_l)^j.
\Eeqar
On the other hand $\et{T}{}^m_l$ and $\ot{T}{}^m_l$ are purely tensorial
and satisfy the following common set of equations:
\Beqar
&&(\Delta_2 + l^2+l-4)(T^m_l)_{jk}=0,\\
&&(T^m_l)^j_j=0,\\
&&D_k(T^m_l)^{jk}=-{1\over2}(l-1)(l+2)(V^m_l)^j.
\Eeqar

Note that $\et{T}{}^m_l$ and $\ot{T}{}^m_l$ are nontrivial only for $l\ge2$
while $\es{T}{}^m_l$ does not vanish for all $l\ge0$.  These three types of
tensor harmonics are mutually orthogonal with respect to the inner product
$\int_{S^2} (\overline{T^m_l})^{jk}(T'{}^m_l)_{jk}$. Hence they are
linearly independent.

\section{Even-Mode Master Variable}

In this appendix we show that the set of equations (\ref{PEeq:e1}) and
Eq.(\ref{PEeq:e2}) is equivalent to the set of equations
(\ref{FbyEpot1}) and (\ref{FbyEpot2}).

First from (\ref{PEeq:e1}) $F^a_b$ in the $(t,r)$ coordinates can be
expressed in terms of two functions $a$ and $b$ as
\Beqar
&& F^t_r=-F^r_t= a-b,\\
&& F^t_t = -F^r_r=-(a+b).
\Eeqar
In terms of $a$, $b$ and the null coordinates $(u,v)$ of $\N$ defined by
\Beq
u=t-r, \qquad v=t+r,
\Eeq
Eq.(\ref{PEeq:e2}) is written as
\Beqar
&& \partial_u F = -4 \partial_v a,\\
&& \partial_v F = -4 \partial_u b.
\Eeqar

If we introduce the quantities $\Phi_1$ and $\Phi_2$ by
\Beq
a=\partial_u^2 \Phi_1, \quad b=\partial_v^2 \Phi_2,
\Eeq
These equations are solved with respect to $F$ as
\Beqar
&&F=-4\partial_u\partial_v\Phi_1 + f_1(v),\\
&&F=-4\partial_u\partial_v\Phi_2 + f_2(u),
\Eeqar
where $f_1(v)$ and $f_2(u)$ are arbitrary functions of $v$ and $u$,
respectively. Eliminating $F$ from these equations and integrating by $u$
and $v$, we obtain
\Beq
4(\Phi_2-\Phi_1)=u\int dv f_1(v)-v\int du f_2(u) +f_3(v)+f_4(u),
\Eeq
where $f_3(v)$ and $f_4(u)$ are arbitrary functions of $v$ and $u$,
respectively.  Hence $a$, $b$ and $F$ are expressed only in terms of the
single function $\epot$ defined by
\Beqar
&4\epot& := 4\Phi_1+u\int dv f_1(v) + f_3(v), \nonumber\\
&&\;=4\Phi_2 + v\int du f_2(u) - f_4(u),
\Eeqar
as
\Beqar
&& a= \partial_u^2 \epot,\\
&& b= \partial_v^2 \epot,\\
&& F= -4\partial_u\partial_v \epot.
\Eeqar
Rewriting these expressions in the covariant form yields
Eqs.(\ref{FbyEpot1}) and (\ref{FbyEpot2}).



\end{document}